\documentclass[aps,prl,twocolumn,superscriptaddress]{revtex4}
\usepackage{graphicx}

\begin{document}

\title{Transmissivity directional hysteresis of a nonlinear metamaterial slab with very small linear permittivity}

\author{A. Ciattoni}
\email{alessandro.ciattoni@aquila.infn.it} \affiliation{Consiglio Nazionale delle Ricerche, CNR-SPIN 67100 L'Aquila, Italy
              and Dipartimento di Fisica, Universit\`{a} dell'Aquila, 67100 L'Aquila, Italy}

\author{C. Rizza}
\affiliation{Dipartimento di Ingegneria Elettrica e dell'Informazione, Universit$\grave{a}$ dell'Aquila, 67100 L'Aquila, Italy}

\author{E. Palange}
\affiliation{Dipartimento di Ingegneria Elettrica e dell'Informazione, Universit$\grave{a}$ dell'Aquila, 67100 L'Aquila, Italy}

\date{\today}

\begin{abstract}
We investigate propagation of a transverse magnetic field through a nonlinear metamaterial slab of sub-wavelength thickness and with a very small and
negative linear dielectric permittivity. We prove that, for a given input intensity, the output intensity is a multi-valued function of the field
incidence angle so that the transmissivity exhibits angular multi-stability and a pronounced directional hysteresis behavior. The predicted
directional hysteresis is a consequence of the fact that the linear and nonlinear contributions to the overall dielectric response can be comparable
so that the electromagnetic matching conditions at the output slab boundary allow more than one field configurations within the slab to be compatible
with the transmitted field.
\end{abstract}


\maketitle
Optical hysteresis behavior and bistability are fascinating nonlinear phenomena which have attracted a large research interest in the last decades
\cite{Chennn} mainly for their potential photonic applications as optical memories, logic gates and optical computing devices \cite{Abraha}. The
possibility of engineering the dielectric permittivity and the magnetic permeability offered by metamaterials \cite{Smith1} has recently allowed to
prove that the feedback mechanism supporting bistability is facilitated by the the opposite directionality of the phase velocity and the energy flow
in the negative index metamaterial \cite{Litchi}. Nonlinear propagation of light through alternating slabs of positive and negative refractive index
materials characterized by a vanishing average refractive index has been considered and bistable switching \cite{Feisee} and gap soliton formation
\cite{Hegdee} have been predicted. Metal-dielectric multilayer structures have also proved to be media exhibiting interesting bistable behavior
\cite{Husako} since the small average dielectric permittivity combined with the nonlinearity allows the sign of the effective overall dielectric
constant to be dependent on the optical intensity \cite{Ciatt1,Ciatt2}.

In this Letter, we investigate electromagnetic transmission through a metamaterial slab of sub-wavelength thickness, characterized by a very small
and negative linear dielectric permittivity and exhibiting focusing Kerr nonlinearity. We numerically solve Maxwell equations for the problem of
reflection and transmission of an inclined incident transverse magnetic (TM) plane wave and we show that the slab transmissivity is, for a given
input intensity, a multi-valued function of the field incidence angle. The novel directional hysteresis behavior is due to the fact that the slab can
host the extreme nonlinear regime where the linear and nonlinear contributions to the overall dielectric response can be comparable \cite{Ciatt1} so
that the electromagnetic matching conditions at the output slab boundary allows more than one possible field configurations compatible with the
transmitted field.

\begin{figure}
\includegraphics[width=0.45\textwidth]{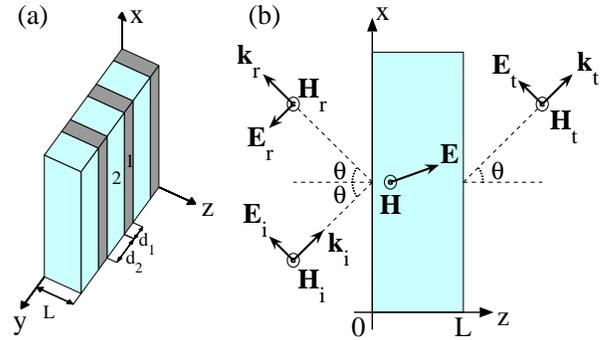}
\caption{(Color online) (a) Geometry of the metamaterial slab. (b) Geometry of the TM incident (i), reflected (r) and transmitted
(t) plane waves together with the TM field within the slab.}
\end{figure}

Consider a transverse magnetic (TM) monochromatic field (with time dependence $\exp(-i \omega t)$) ${\bf E} = E_x(x,z) \hat{\bf e}_x + E_z(x,z)
\hat{\bf e}_z$ and ${\bf H} = H_y(x,z) \hat{\bf e}_y$ propagating through the metamaterial slab, embedded in vacuum, reported in Fig.1(a) of
sub-wavelength thickness (along the $z$-axis) $L = 1/k_0$ ($k_0 = \omega /c$) and consisting of alternating, along the $y$-axis, two nonlinear
isotropic and non-magnetic Kerr layers of thicknesses $d_1$ and $d_2$. If the spatial period is much smaller than the vacuum wavelength (i.e. $d_1 +
d_2 \ll 2 \pi / k_0$), the slab behaves as an homogeneous medium characterized by the dielectric response (constitutive relations)
\begin{eqnarray} \label{constituent}
D_x = \epsilon_0 \left\{ \epsilon E_x + \chi \left[\left(|E_x|^2+|E_z|^2 \right) E_x + \frac{1}{2} \left(E_x^2+E_z^2 \right) E_x^* \right] \right\}, \nonumber \\
D_z = \epsilon_0 \left\{ \epsilon E_z + \chi \left[\left(|E_x|^2+|E_z|^2 \right) E_z + \frac{1}{2} \left(E_x^2+E_z^2 \right) E_z^* \right] \right\},
\end{eqnarray}
where the effective dielectric permittivity $\epsilon = f \epsilon_1 + (1-f) \epsilon_2$ and effective nonlinear susceptibility $\chi = f \chi_1 +
(1-f) \chi_2$ are the averages of their layers counterparts (here $f= d_1/(d_1+d_2)$ is the volume filling fraction of layer $1$) \cite{Ciatt1}.
Therefore, employing both nonlinear negative and positive dielectrics ($Re(\epsilon_1) <0$ and $Re(\epsilon_2) >0$), conditions can be found so that
the effective dielectric permittivity $\epsilon$ has a negative real part (with a very small imaginary part) and very close to zero, and $\chi>0$
(see below for an explicit example) which is the situation we focus on in this Letter. Note that, due the the small value of $\epsilon$, the
effective medium is able to support the extreme nonlinear regime where the nonlinear contributions to Eqs.(\ref{constituent}) can be comparable of
even greater than the linear part $\epsilon {\bf E}$ \cite{Ciatt1,Ciatt2}.

A TM plane wave $({\bf E}_i,{\bf H}_i)$ incoming from vacuum ($z<0$) and impinging on the slab at $z=0$ with incidence angle $\theta$, produces a TM
reflected plane wave $({\bf E}_r,{\bf H}_r)$ for $z<0$ and a TM transmitted plane wave $({\bf E}_t,{\bf H}_t)$ for $z>L$, as reported in Fig.1(b).
The fields of the three waves are ${\bf E}_s = E_s (\hat{\bf e}_y \times {\bf k}_s/k_0) \exp \left(i {\bf k}_s \cdot {\bf r}\right)$, ${\bf H}_s =
E_s \sqrt{\epsilon_0/\mu_0} \hat{\bf e}_y \exp \left(i {\bf k}_s \cdot {\bf r}\right)$ where $s=(i,r,t)$, $E_i$, $E_r$ and $E_t$ are the amplitudes
of the three waves and ${\bf k}_i={\bf k}_t= k_0 \left(\sin \theta \hat{\bf e}_x + \cos \theta \hat{\bf e}_z \right)$, ${\bf k}_r= k_0 \left(\sin
\theta \hat{\bf e}_x - \cos \theta \hat{\bf e}_z \right)$ are the wave vectors of the three waves. The TM electromagnetic field within the slab is of
the form ${\bf E} = \left[ E_x(z) \hat{\bf e}_x + E_z(z) \hat{\bf e}_z \right] \exp(i k_0 x \sin \theta )$, ${\bf H} = \left[ H_y(z) \hat{\bf e}_y
\right] \exp(i k_0 x \sin \theta )$ so that, Maxwell equations $\nabla \times {\bf E} = i \omega \mu_0 {\bf H}$, $\nabla \times {\bf H} = -i \omega
{\bf D}$  yields the system
\begin{eqnarray} \label{Maxwell}
\frac{d E_x}{dz} &=& i (k_0 \sin \theta) E_z + i \omega \mu_0 H_y, \nonumber \\
\frac{d H_y}{dz} &=& i \omega D_x, \nonumber \\
(k_0 \sin \theta) H_y &=& - \omega D_z,
\end{eqnarray}
where $D_x$ and $D_z$ are given by Eqs.(\ref{constituent}) and where the third equation implicitly defines $E_z$ as a function of $E_x$ and $H_y$. In
order to evaluate the slab transmissivity $T=|E_t|^2/|E_i|^2$, we specify the amplitude of the transmitted field $E_t$ at the output face $z=L$ and
successively solve Eqs.(\ref{Maxwell}) all the way to the input face at $z=0$. At the slab input and output faces we require the continuity of the
tangential components of the electric and magnetic fields, namely $(E_i - E_r) \cos \theta = E_x(0)$, $\sqrt{\epsilon_0 / \mu_0} (E_i + E_r) =
H_y(0)$ (at $z=0$) and $E_x(L) = E_t \cos \theta \exp(i \cos\theta)$, $H_y (L) = \sqrt{\epsilon_0 / \mu_0}  E_t \exp(i \cos\theta)$ (at $z=L$) where
use has been made of the fact that $L =1/k_0$. In order to evaluate the transmissivity for a given input intensity as a function of $\theta$ we have
first evaluated a number of amplitudes $E_i$ for various $E_t$ and $\theta$ and subsequently collected all the configurations corresponding to the
same $E_i$. The resulting slab transmissivity $T$ as a function of $\theta$ is reported in Fig.2 for $|E_i| = 10^{-6} |\chi|^{-1/2}$ (dashed line)
and for $|E_i| = 5.3 \cdot 10^{-4} |\chi|^{-1/2}$ (solid line), for a slab with $\epsilon = -0.001$ and $\chi>0$. The dashed line transmissivity is
evaluated for an input intensity so low that it coincides with the transmissivity of the linear slab and it has been reported for comparison
purposes. Note that the solid line transmissivity is a multi-valued function of the incident angle since it is threefold in the angular range $-0.028
< \theta < 0.028$ (radians).

\begin{figure}
\includegraphics[width=0.45\textwidth]{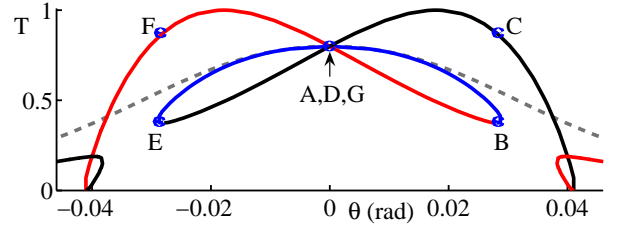}
\caption{(Color online) Slab transmissivity $T$ as a function of $\theta$ for $|E_i| = 10^{-6} |\chi|^{-1/2}$ (dashed line) and
for $|E_i| = 5.3 \cdot 10^{-4} |\chi|^{-1/2}$ (solid line), for a slab with $\epsilon = -0.001$ and $\chi>0$.}
\end{figure}

In order to physically grasp the predicted novel kind of multi-stability it is convenient to consider the field matching conditions at the slab
boundaries. Exploiting the third of Eqs.(\ref{Maxwell}) it is evident that the continuity of $H_y$ across the slab boundaries is equivalent to the
continuity of $D_z$, so that after setting $E_z(L) = A_z ( E_t/|E_t| ) \exp(i \cos \theta)$ (where the amplitude $A_z$ is real) and exploiting the
second of Eqs.(\ref{constituent}) together with the continuity of $E_x$, the matching condition $D_z(L^-)=D_z(L^+)$ yields
\begin{equation} \label{Ezmatch}
\left[\epsilon  + \frac{3}{2} \chi \left( |E_t|^2 \cos^2 \theta + A_z^2 \right) \right] A_z+ |E_t| \sin \theta = 0,
\end{equation}
which is a relation that, for a given $E_t$ and $\theta$ yields the possible values of $A_z$. Note that, in the linear regime ($\chi=0$)
Eq.(\ref{Ezmatch}) yields $A_z = - |E_t| \sin \theta / \epsilon$ i.e. the boundary conditions allow only one possible value of $A_z$. The possible
values of $|\chi|^{1/2} A_z$ obtained by Eq.(\ref{Ezmatch}), in the case $\epsilon = -0.001$ and $\chi>0$, are plotted in Fig.3 from which we note
that the surface is folded so that there are regions where, for a given pair $|\chi|^{1/2}|E_t|$ and $\theta$, there are more than one accessible
value of $|\chi|^{1/2} A_z$. In other words, there are more than one possible electromagnetic fields compatible with the same output field and
incidence angle. The origin of the surface folding is easily discussed by considering the situation of vanishing incidence angle ($\theta=0$) for
which Eq.(\ref{Ezmatch}), in addition to the standard solution $A_z^{(0)}=0$, also admits the solutions $|\chi|^{1/2} A_z^{(\pm)} = \pm \sqrt{-
\textrm{sign} (\chi)\left(2\epsilon/3 + \chi |E_t|^2 \right)}$ which are reported with two dotted lines in Fig.3 on the upper and lower surface
sheets. The surface is folded if $|\chi|^{1/2} A_z^{(+)}$ is real (existence of the upper and lower surface sheets) and therefore a necessary
condition is $\textrm{sign}(\epsilon \chi) = -1$. It is worth noting that $|\chi|^{1/2} A_z^{(\pm)}$ are obtained by the vanishing of the term in
square brackets of Eqs.(\ref{Ezmatch}) (for $\theta =0$) and this corresponds to requiring the nonlinear contribution to exactly compensate the
linear part in the second of Eqs.(\ref{constituent}). Therefore, as opposite to the situation we are considering in this Letter where $|\epsilon| \ll
1$, in standard materials where $\epsilon$ is of the order of unity, the required nonlinearity would be so large to effectively forbid the folding of
the surface (and the consequent feasibility of the directional multi-stability). Note now that each of the point of Fig.2 corresponds to a field
configuration satisfying the matching conditions so that the curve of Fig.2 has an image on the surface of Fig.3 which we report as a solid line. The
above mentioned multi-valuedness of the slab transmissivity of Fig.2 is therefore explained since the different values of transmissivity at a give
$\theta$ corresponds to different fields with the same input intensity and having different output longitudinal component $A_z$ belonging to
different sheets of the folded surface.

\begin{figure}
\includegraphics[width=0.45\textwidth]{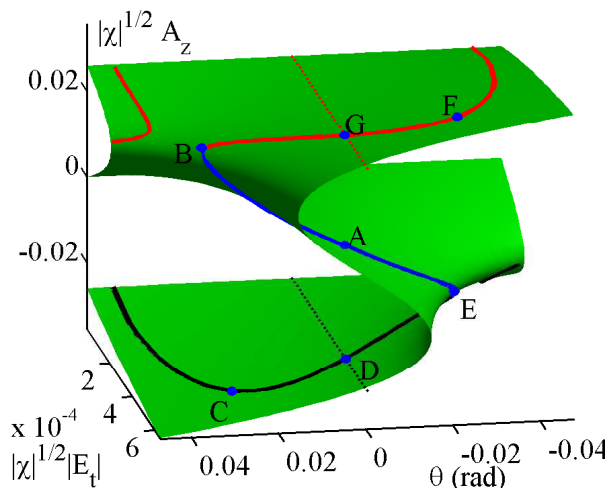}
\caption{(Color online) Surface $|\chi|^{1/2} A_z$ obtained by Eq.(\ref{Ezmatch}), in the case $\epsilon = -0.001$ and $\chi>0$.}
\end{figure}

The multi-valuedness of the slab transmissivity is accompanied by a directional hysteresis behavior. Suppose the incident plane wave to be initially
normally impinging on the slab ($\theta =0$) so that the state $A$ (with $A_z=0$) of Fig.(3) and Fig.(2) is excited since before the illumination the
slab was not polarized. By increasing the incidence angle $\theta$, the electromagnetic state continuously varies toward the state $B$ (see Fig.3)
and the transmissivity becomes to depart from its linear behavior (see Fig.2) since $A_z$ arrives at the upper sheet of the surface where its value
is greater than its linear counterpart. From the state B, if $\theta$ is further increased, from Fig.3 it is evident that the state has to undergo a
sudden jump to the state C (belonging to the lower sheet of the surface) since there is no allowed state on the upper sheet which is continuously
joined with B at a greater $\theta$. Correspondingly, the transmissivity undergoes a sudden jump to a higher value ($T(C)>T(B)$, see Fig.2) since
$A_z$ has a large magnitude on the lower surface sheet. Suppose now that, from the state C, $\theta$ is decreased so that the electromagnetic state
now continuously varies toward the state $D$ on the lower sheet (see Fig.3). Correspondingly, the transmissivity varies along the curve from C to D
of Fig.2 assuming values different from those of the forward path from A to B, i.e. directional hysteresis behavior occurs. Evidently, the procedure
can be extended to the states E-F-G-B and so on. Note that the three states A, D and G, while having different values of $A_z$ (see Fig.3), share a
common value of the transmissivity (see Fig.2) since they have $\theta=0$ so that, from the first of Eqs.(\ref{Maxwell}), $E_z$ can not affect
electromagnetic propagation.

As a specific example, consider a TM field of wavelength $\lambda = 810 \: nm$ so that the slab thickness is $L=1/k_0 = \lambda/2\pi = 129 \: nm$. We
choose the slab layers of type $1$ and $2$ (see Fig.1(a)) to be filled with silver (1) and AlGaAs (2), respectively, so that $\epsilon_1 =
-28.81+1.55i$, $\chi_1 = 3.46 \times 10^{-16} \: m^2/V^2$ and $\epsilon_2=10.9-0.5865i$, $\chi_2 = 6.56  \times 10^{-18} \: m^2/V^2$. Note that we
are exploiting the very large silver nonlinear susceptibility \cite{Husako} and the fact that AlGaAs optically amplifies the radiation at the chosen
wavelength if the sample is pumped by ultra-violet light so that the above negative value of $Im(\epsilon_2)$ can be attained by adjusting the pump
laser intensity \cite{Ramakr}. Choosing $d_1 = 14 \: nm$ and $d_2 = 37 \: nm$, we obtain $\epsilon = -0.001-0.00001i$ and $\chi = 10^{-16} \:
m^2/V^2$ which are values adequate to observe the directional hysteresis. The intensity of the incident plane wave is $I=(1/2)\sqrt{\epsilon_0 /
\mu_0} |E_i|^2$ which, for the amplitude $E_i$ chosen for evaluating the transmissivity of Fig.2 and the above effective nonlinear susceptibility
$\chi$, yields the value $I = 372 \: W / cm^2$, an extremely small value if compared to those ($\sim MW/cm^2$) normally required for observing the
standard optical bistability.

In conclusion, we have shown that a sub-wavelength nonlinear metamaterial slab with a very small linear permittivity is characterized by a
transmissivity exhibiting a novel kind of directional hysteresis behavior which is observable at very low intensities of the incident field.

\end{document}